\documentclass[aps,english,showpacs,twocolumn]{revtex4-1}
\usepackage{amsfonts}
\usepackage{amssymb}
\usepackage{amsmath}
\usepackage{graphicx}
\usepackage{epsfig}
\usepackage{color}
\usepackage{amsmath,bm}
\usepackage{booktabs}
\usepackage[colorlinks=true, linkcolor=blue, citecolor=blue, urlcolor=blue]{hyperref}

\begin{document}

\title{Exceptional-point-induced dynamic sensitivity to particle-number
parity}
\author{J. Y. Liu-Sun}
\author{Z. Song}
\email{songtc@nankai.edu.cn}

\begin{abstract}
As an exclusive feature of a non-Hermitian system, the existence of
exceptional points (EPs) depends not only on the details of the Hamiltonian
but also on the particle-number filling and the particle statistics. In this
paper, we study many-particle EPs in a Bose Hubbard chain with two end-site
resonant imaginary potentials. Starting from a single-particle coalescing
eigenstate, we construct $n$-particle condensate eigenstates for the cases
with zero and infinite $U$. Compared with the free bosonic case, where the $%
n $-particle condensate eigenstate is an $(n+1)$-th-order coalescing state,
the hardcore-boson counterpart is a second-order coalescing state for odd $n$%
, while it is not for even $n$. The difference in particle-number parity
results in distinct quenching dynamics of the condensate states,
highlighting the role of parity in system behavior. Our finding may
stimulate research on the dynamic sensitivity to particle-number parity.
\end{abstract}

\affiliation{School of Physics, Nankai University, Tianjin 300071, China}
\maketitle


\section{Introduction}

\label{Introduction}

The theoretical \cite{bender2007making,moiseyev2011non,krasnok2019anomalies}
and experimental studies \cite%
{guo2009observation,ruter2010observation,peng2014parity,feng2014single,hodaei2014parity,feng2017non,longhi2018parity,el2018non,miri2019,ozdemir2019parity,wu2019observation}
on the non-Hermitian system\ indicate that the interplay of lattice geometry
and non-Hermitian elements, such as imaginary on-site potential \cite%
{jin2013scaling,jin2011hermitian} and asymmetry tunneling \cite%
{XZZhangAP2013} can induce exotic single-particle dynamics \cite%
{PRL3,longhi2017unidirectional,XQLPRA2015,LJinPRL2018,LJinCPL2021}, which
never happens in a Hermitian system. Apparently, for an interacting system,
the particle-number filling and the particle statistics are also crucial, as
they determine the features of the non-Hermitian Hamiltonian matrix. As an
exclusive feature of a non-Hermitian system, the existence of exceptional
points (EPs) occurs when two or more eigenstates coalesce, and usually
associates with the non-Hermitian phase transition \cite%
{Keldysh,Kato,moiseyev2011non,Berry2004,Heiss2012,miri2019}. In the vicinity
of the exception point, the system supports a anomalous dynamical behavior.
Then, the EP plays a pivotal role in both intriguing dynamics and
applications, including asymmetric mode switching \cite{Doppler2016},
unidirectional lasing \cite{Ramezani2014,Peng2016,LJinPRL2018}, and enhanced
optical sensing \cite%
{Wiersig2014,Liu2016,Hodaei2017,Chen2017,Lau2018,Zhang2019,Lai2019,Hokmabadi2019}%
. Recently, EP dynamics is employed to engineer a target quantum state \cite%
{TEL2014,CL2015,XMY2020,ZXZ2020} and probe quantum phase transitions \cite%
{ZKLprl}.

In this paper, we focus on the many-particle system, in which EPs are
induced by the particle number filling. Our study is based on the exact
result for the Bose-Hubbard chain with two end-site potentials and resonant
imaginary potentials. A set of $n$-particle condensate eigenstates are
constructed in the cases with zero and infinite $U$. In particular, we will
show that such an eigenstate is a second-order coalescing state for odd $n$,
while it is not for even $n$. It has been demonstrated that the dynamics of
the system with parameters far away from, near, and at the EP exhibit
extremely different behaviors \cite%
{YXM2021Quantum,YXM2022Dynamic,ZHANG2025170049}. Therefore, it is expected
that the difference in particle-number parity should result in distinct
quenching dynamics of the condensate states, highlighting the role of parity
in system behavior. Numerical simulations are performed to demonstrate our
predictions. In quantum information, particle-number parity is a fundamental
quantum property, which can serve as a quantum error-correction mechanism or
as a way to classify and reconstruct quantum states \cite{Besse2020Parity}. Our finding may stimulate
research on the dynamic sensitivity to particle-number parity.

This paper is organized as follows. In Section \ref{Model Hamiltonians}, we
present the model Hamiltonian and analyze the symmetries of the system. In
Sections \ref{Condensate eigenstates} and \ref{Many-particle coalescing
states}, we construct the condensate eigenstates and investigate\ their
properties associated with the particle filling, respectively. Section \ref%
{Parity-dependent dynamics} is devoted to the numerical simulation of the
model, revealing the dynamical signature of the particle-number parity.\
Finally, we give a summary and discussion in Section \ref{Summary}. Some
details of the derivations are provided in the Appendix.

\begin{figure*}[t]
	\centering
	\includegraphics[width=1.0\textwidth]{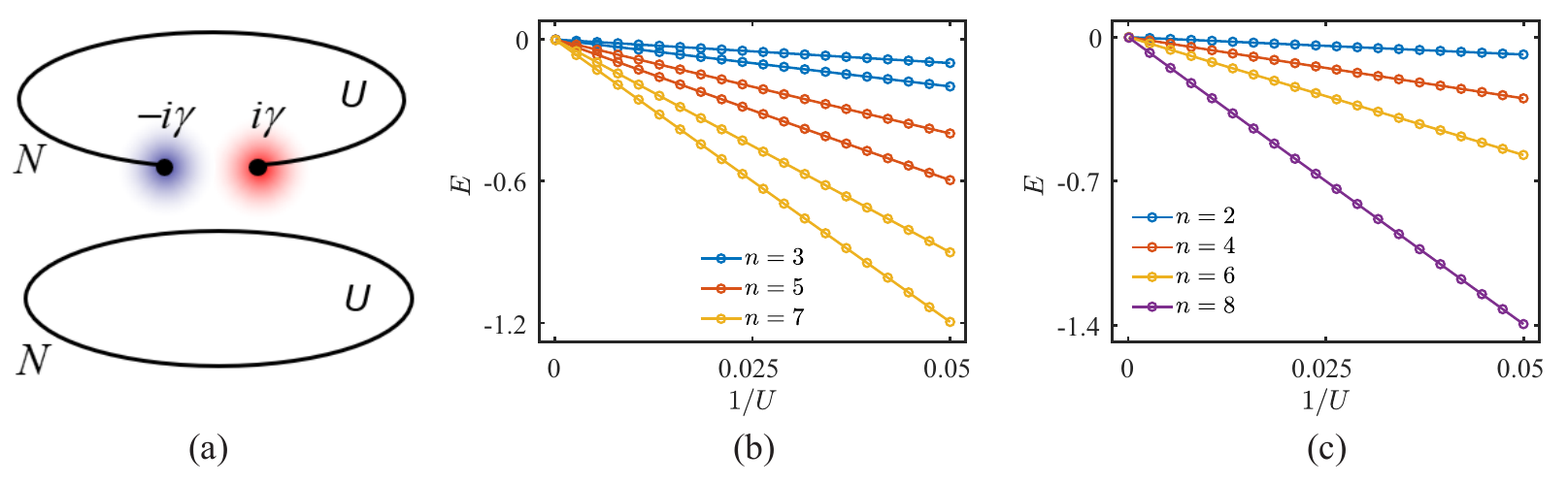}
	\caption{(a) Schematic illustration of the systems related to this work. The
		upper panel depicts a non-Hermitian $\mathcal{PT}$-symmetric $N$-site
		Bose-Hubbard chain with two end-site imaginary potentials. The on-site
		Hubbard interaction strength is $U$, measured in units of the
		nearest-neighbor hopping strength. The lower panel shows a Hermitian $N$%
		-site Bose-Hubbard ring with the same $U$. One of the main conclusions of
		this work is that both systems share the same many-particle condensate
		eigenstates when $\protect\gamma =1$ and $U=0$ or $U=\infty $ (b) Plots of
		energy levels as functions of $U$ approaching the condensate states $%
		\left\vert \protect\psi _{n}\right\rangle $, given by Eq. (\protect\ref%
		{condensate state}), of the Hamiltonian given by Eq. (\protect\ref{H_U}%
		) with odd filling number $n$. The results are obtained by exact
		diagonalizations for 8-site systems. (c) The same plots for the cases with
		even filling number $n$. We find that there are two levels that approach
		zero in each case with odd $n$, while only a single level approaches zero in
		each case with even $n$. This indicates that $\left\vert \protect\psi %
		_{n}\right\rangle $ is a coalescing state at a 2nd-order EP for odd $n$.}
	\label{fig1}
\end{figure*}

\section{Model Hamiltonians}

\label{Model Hamiltonians}

The Bose-Hubbard model plays an important role in quantum many-body physics,
since it embodies essential features of ultracold atoms in optical lattices 
\cite{Greiner,Jaksch}. Recently, a non-Hermitian Bose-Hubbard model with an
imaginary potential on the edge has been investigated \cite%
{Hiller,Graefe08PRL,Graefe08PRA}.

The Hamiltonian of a $\mathcal{PT}$-symmetric non-Hermitian Bose-Hubbard
reads

\begin{equation}
H=-\sum_{l=1}^{N-1}\left( b_{l}^{\dag }b_{l+1}+\text{H.c.}\right) +\frac{U}{2%
}\sum_{l=1}^{N}b_{l}^{\dag 2}b_{l}^{2}+i\gamma \left( n_{1}-n_{N}\right) ,
\label{H_U}
\end{equation}%
where $b_{l}^{\dag }$ is the creation operator of the boson at the $l$th
site and the tunneling strength sets the energy scale. The nonlinear
coupling strength and the on-site potential are denoted by $U$ and $i\gamma $%
. In Fig. \ref{fig1}(a), the above Hamiltonian is schematically illustrated.
It is a $\mathcal{PT}$-symmetric model, i.e., $[\mathcal{PT},H]=0$, where
the action of the parity operator $\mathcal{P}$ is defined by $\mathcal{P}%
:l\rightarrow N+1-l$ and the time-reversal operator $\mathcal{T}$ by $%
\mathcal{T}:i\rightarrow -i$. In a non-Hermitian system, although the
particle probability is no long conservative, the particle number $\hat{N}%
=\sum_{l=1}^{N}b_{l}^{\dag }b_{l}$ still shares the common eigenfunctions
with the Hamiltonian due to the commutation relation $[\hat{N},H]=0$. To
date, no exact solution exists for this Hamiltonian with given $U$ and $%
\gamma $. In the following, we will present some exact results for the
Hamiltonian with infinite $U$ and $\gamma =1$. We begin with the
non-interacting Hamiltonian $H_{\text{FB}}=H(U=0)$, whose solutions can be
constructed from the single-particle solution of $H$. In the single-particle
invariant subspace, it turns out that the EP occurs at the critical value $%
\gamma _{c}$, where $\gamma _{c}=\pm 1$ ($\pm \sqrt{\left( l+1\right) /l}$, $%
l=1,2,...$), for the case of $N=2l$ ($N=2l+1$) \cite{JLPT}. The key starting
point of this paper is that, for $N=4l$, the coalescing eigenstate takes the
simple form%
\begin{equation}
\left\vert \phi _{\mathrm{c}}\right\rangle =\left(\sum_{j=1}^{N}e^{i\frac{\pi }{2}%
j}b_{j}^{\dagger }\right)\left\vert \text{vac}\right\rangle ,
\end{equation}%
where $\left\vert \text{vac}\right\rangle $\ is the bosonic vacuum state.
Note that although the Hamiltonian $H_{\text{FB}}$\ lacks translational
symmetry,

\begin{equation}
T_{1}H_{\mathrm{FB}}T_{1}^{-1}\neq H_{\mathrm{FB}},
\end{equation}%
its coalescing eigenstate $\left\vert \phi _{\mathrm{c}}\right\rangle $
satisfies%
\begin{equation}
T_{1}\left\vert \phi _{\mathrm{c}}\right\rangle =e^{-i\frac{\pi }{2}%
}\left\vert \phi _{\mathrm{c}}\right\rangle .
\end{equation}%
Here, $T_{1}$\ is translational operator defined as 
\begin{equation}
T_{1}b_{j}T_{1}^{-1}=b_{j+1},
\end{equation}%
with the periodic boundary condition $b_{N+1}=b_{1}$.

\section{Condensate eigenstates}

\label{Condensate eigenstates}

In this section, we focus on the Hamiltonian $H$ in Eq. (\ref{H_U}) under
resonant conditions, taking $U=\infty $, $N=4l$ ($l=1,2,\ldots $) and $%
\gamma =1$. The investigation of this model is based on the result of
single-particle for $H_{\text{FB}}$. This hardcore-boson Hamiltonian is
expressed as%
\begin{equation}
H_{\mathrm{HB}}=-\sum_{j=1}^{N-1}a_{j}^{\dagger }a_{j+1}+\text{\textrm{H.c.}}%
+i\left( a_{1}^{\dagger }a_{1}-a_{N}^{\dagger }a_{N}\right) ,
\end{equation}%
where $a_{l}^{\dagger }$ is the hardcore-boson creation operator at the
position $l$, and $n_{l}=a_{l}^{\dagger }a_{l}$. The hardcore-boson
operators satisfy the commutation relations%
\begin{equation}
\left\{ a_{l},a_{l}^{\dagger }\right\} =1,\left\{ a_{l},a_{l}\right\} =0,
\end{equation}%
and%
\begin{equation}
\left[ a_{j},a_{l}^{\dagger }\right] =0,\left[ a_{j},a_{l}\right] =0,
\end{equation}%
for $j\neq l$. The total particle number operator, $n=\sum_{l}n_{l}$, is a
conserved quantity because it commutes with the Hamiltonian. Then one can
still investigate the system in each invariant subspace with fixed particle
number $n$. The model can be mapped to the spin-$1/2$\ XXZ model \cite%
{Matsubara1956Alatticemodel}, which enables the application of our results
to both hardcore-boson and quantum spin systems.

We will construct the eigenstates based on the single-particle eigenstate $%
\left\vert \phi _{\mathrm{c}}\right\rangle $. To this end, we introduce a
set of operators%
\begin{eqnarray}
\eta _{\frac{\pi }{2}}^{+} &=&\left( \eta _{\frac{\pi }{2}}^{-}\right)
^{\dagger}=\sum_{j=1}^{N}e^{i\frac{\pi }{2}j}a_{j}^{\dagger }, \\
\eta _{\frac{\pi }{2}}^{z} &=&\frac{1}{2}\sum_{j=1}^{N}\left( a_{j}^{\dagger
}a_{j}-1\right) ,
\end{eqnarray}%
which are pseudo-spin\ operators, satisfying the su(2) algebra, $\left[ \eta
_{\frac{\pi }{2}}^{+},\eta _{\frac{\pi }{2}}^{-}\right] =2\eta _{\frac{\pi }{%
2}}^{z}$\ and $\left[ \eta _{\frac{\pi }{2}}^{z},\eta _{\frac{\pi }{2}}^{\pm
}\right] =\pm \eta _{\frac{\pi }{2}}^{\pm }$. On the other hand, a
straightforward derivations show that%
\begin{equation}
\left[ \eta _{\frac{\pi }{2}}^{+},\left[ \eta _{\frac{\pi }{2}}^{+},H_{%
\mathrm{HB}}\right] \right] =0,
\end{equation}%
and%
\begin{equation}
\left[ \eta _{\frac{\pi }{2}}^{+},H_{\mathrm{HB}}\right] \left\vert \text{vac%
}\right\rangle =0,
\end{equation}%
and%
\begin{equation}
H_{\mathrm{HB}}\left\vert \text{vac}\right\rangle =0,
\end{equation}%
which meet the conditions of the the restricted spectrum generating algebra
(RSGA) formalism introduced in Ref. \cite{Moudgalya2020Hubbard}. Here, $%
\left\vert \text{vac}\right\rangle $\ denotes the vacuum state of both
bosons and fermions.\ We therefore conclude that a set of eigenstates of $H_{%
\mathrm{HB}}$ can be written as%
\begin{equation}
\left\vert \psi _{m}\right\rangle =\frac{1}{m!\sqrt{C_{N}^{m}}}(\eta _{\frac{%
\pi }{2}}^{+})^{m}\left\vert \text{vac}\right\rangle ,
\label{condensate state}
\end{equation}%
which satisfy $H_{\mathrm{HB}}\left\vert \psi _{m}\right\rangle =0$ for $%
m\in \left[ 0,N\right] $.

In addition, the state $\left\vert \psi _{n}\right\rangle $\ possesses ODLRO
due to the fact that the correlation function 
\begin{equation}
\left\langle \psi _{n}\right\vert a_{p}^{\dagger }a_{q}\left\vert \psi
_{n}\right\rangle =i^{q-p}\frac{n(N-n)}{N(N-1)},  \label{aa}
\end{equation}%
does not decay as $\left\vert q-p\right\vert $\ increases (see Ref. \cite%
{yang1962concept}). Indeed, $\left\vert \psi _{n}\right\rangle $ can be
written as%
\begin{eqnarray}
&&\left\vert \psi _{n}(N)\right\rangle =\frac{1}{\sqrt{C_{N}^{n}}}%
\left(\sum_{\{l_{j}\}}i^{\sum_{j=1}^{n}l_{j}}\prod\limits_{j=1}^{n}a_{l_{j}}^{%
\dagger }\right)\left\vert \text{\textrm{vac}}\right\rangle  \notag \\
&=&\frac{1}{\sqrt{C_{N}^{n}}} \left[ (\sum_{\{l_{j}\}\backslash
	\{p\}}i^{\sum\limits_{j=1}^{n-1}l_{j}}\prod\limits_{j=1}^{n-1}a_{l_{j}}^{%
	\dagger })i^{p}a_{p}^{\dagger } \right. \notag \\
&& \left. +(\sum_{\{l_{j}\}\backslash
	\{p\}}i^{\sum\limits_{j=1}^{n}l_{j}}\prod\limits_{j=1}^{n}a_{l_{j}}^{%
	\dagger }) \right] \left\vert \text{\textrm{vac}}\right\rangle ,
\end{eqnarray}%
where $\{l_{j}\}\equiv \{l_{1},...,l_{j},...l_{n}\}$\ denotes a combination,
or a selection of $n$ distinct numbers from the set $[1,N]$, and $%
\{l_{j}\}\backslash \left\{ p\right\} $\ denotes set difference $%
\{l_{j}\}-\left\{ p\right\} $, representing a set with $n-1$ distinct
numbers in $\{l_{j}\}$\ but not in set $\left\{ p\right\} $. Here, we
express $\left\vert \psi _{n}\right\rangle $ as $\left\vert \psi
_{n}(N)\right\rangle $ without losing the generality of the expression of
the state. The summation $\sum_{\{l_{j}\}}$ ($\sum_{\{l_{j}\}\backslash
\left\{ p\right\} }$) denotes the sum over all combinations in the set $%
\{l_{j}\}$\ ($\{l_{j}\}\backslash \left\{ p\right\} $). A direct derivation
shows that

\begin{eqnarray}
&&a_{q}\left\vert \psi _{n}(N)\right\rangle =\sqrt{\frac{C_{N-2}^{n-2}}{%
C_{N}^{n}}}i^{p+q}a_{p}^{\dagger }\left\vert 0\right\rangle _{p}\left\vert
0\right\rangle _{q}\left\vert \psi _{n-2}(N-2)\right\rangle  \notag \\
&&+\sqrt{\frac{C_{N-2}^{n-1}}{C_{N}^{n}}}i^{q}\left\vert 0\right\rangle
_{p}\left\vert 0\right\rangle _{q}\left\vert \psi _{n-1}(N-2)\right\rangle ,
\end{eqnarray}%
where $\left\vert 0\right\rangle _{j}$ denotes the vacuum state of the
operator $a_{j}$. Applying it to the expression $a_{q}\left\vert \psi
_{n}(N)\right\rangle $ gives the correlation function in Eq. (\ref{aa}),
i.e.,

\begin{eqnarray}
&&\left\langle \psi _{n}(N)\right\vert a_{p}^{\dagger }a_{q}\left\vert \psi
_{n}(N)\right\rangle =i^{q-p}\frac{C_{N-2}^{n-1}}{C_{N}^{n}}  \notag \\
&=&i^{q-p}\frac{n(N-n)}{N(N-1)}.
\end{eqnarray}%
Alternative proofs for similar issues can be found in the appendix of Ref. \cite{Zhang2025Coalescing}.

\section{Many-particle coalescing states}

\label{Many-particle coalescing states}

According to non-Hermitian quantum mechanics, the eigenstates of $H_{\mathrm{%
HB}}$\ and $H_{\mathrm{HB}}^{\dag }$\ form a biorthonormal complete set,
except at the EP. In parallel, the eigenstates of $H_{\mathrm{HB}}^{\dag }$\
can be expressed in the form%
\begin{equation}
\left\vert \varphi _{m}\right\rangle =\frac{1}{m!\sqrt{C_{N}^{m}}}(\eta _{-%
\frac{\pi }{2}}^{+})^{m}\left\vert 0\right\rangle .
\end{equation}%
Importantly, states $\left\vert \psi _{m}\right\rangle $\ and $\left\vert
\varphi _{n}\right\rangle $\ are the coalescing states of $H_{\mathrm{HB}}$\
and $H_{\mathrm{HB}}^{\dag }$, respectively, provided that%
\begin{equation}
\langle \varphi _{n}\left\vert \psi _{n}\right\rangle =0.
\end{equation}%
This allows us to identify the coalescing state. Indeed,

\begin{eqnarray}
\langle \varphi _{n}\left\vert \psi _{n}\right\rangle &=&\left\langle
\varphi _{n}|T_{1}^{-1}T_{1}|\psi _{n}\right\rangle =\left( T_{1}\left\vert
\varphi _{n}\right\rangle \right) ^{\dag }T_{1}\left\vert \psi
_{n}\right\rangle  \notag \\
&=&\left( -1\right) ^{n}\langle \varphi _{n}\left\vert \psi
_{n}\right\rangle ,
\end{eqnarray}%
which shows that $\left\vert \psi _{n}\right\rangle $ is the coalescing
state when the boson number $n$ is odd. Here we would like to point out that
the above proof cannot conclude that $\left\vert \psi _{n}\right\rangle $ is
not the coalescing state when the boson number $n$ is even. In the Appendix %
\ref{Appendix}, a direct derivation shows that%
\begin{equation}
\left\langle \varphi _{n}|\psi _{n}\right\rangle =\left\{ 
\begin{array}{cc}
0, & (n\in \text{\textrm{odd}}) \\ 
\text{\textrm{Nonzero,}} & (n\in \text{\textrm{even}})%
\end{array}%
\right. ,  \label{bi norm}
\end{equation}%
which indicates that $\left\vert \psi _{n}\right\rangle $ is not a
coalescing state for even $n$. It is the central result of this work. In
Fig. \ref{fig1}(b) and (c), we demonstrate this result by numerical
simulations for a finite-size system with finite values of $U$. It indicates
that $\left\vert \psi _{n}\right\rangle $\ is the coalescing state of
2nd-order EP for odd $n$. However, it cannot be proved exactly. Although the
underlying mechanism is not clear, particle statistics should play an
important role. This can be seen from the situation where the particles are
pure bosons and fermions, respectively.

First, consider the free-boson Hamiltonian%
\begin{equation}
H_{\text{FB}}=-\sum_{j=1}^{N-1}b_{j}^{\dagger }b_{j+1}+\text{\textrm{H.c.}}%
+i\left( b_{1}^{\dagger }b_{1}-b_{N}^{\dagger }b_{N}\right) ,
\end{equation}%
where the condensate state%
\begin{equation}
\left\vert \phi _{n}\right\rangle =\frac{1}{\sqrt{n!N^{n}}}\left( %
\sum_{j=1}^{N}e^{i\frac{\pi }{2}j}b_{j}^{\dagger }\right) ^{n}\left\vert 
\text{vac}\right\rangle ,
\end{equation}%
is an $\left( n+1\right) $-th order coalescing state, satisfying $H_{\mathrm{%
FB}}\left\vert \phi _{n}\right\rangle =0$\ for any $n\in \lbrack 0,\infty )$%
. Second, consider the free-fermion Hamiltonian%
\begin{equation}
H_{\text{FF}}=-\sum_{j=1}^{N-1}f_{j}^{\dagger }f_{j+1}+\text{\textrm{H.c.}}%
+i\left( f_{1}^{\dagger }f_{1}-f_{N}^{\dagger }f_{N}\right) ,
\end{equation}%
where the single-fermion eigenstate%
\begin{equation}
\left\vert \phi \right\rangle =\frac{1}{\sqrt{N}}\left( \sum_{j=1}^{N}e^{i%
\frac{\pi }{2}j}f_{j}^{\dagger }\right) \left\vert \text{vac}\right\rangle ,
\end{equation}%
is a 2nd-order coalescing state. Therefore, for any given $n$, all the
coalescing states are second-order. It is clear that the three types of
particles---bosons, hardcore bosons, and fermions---have different types of
EPs.

\begin{figure*}[tbph]
	\centering
	\includegraphics[width=1.0\textwidth]{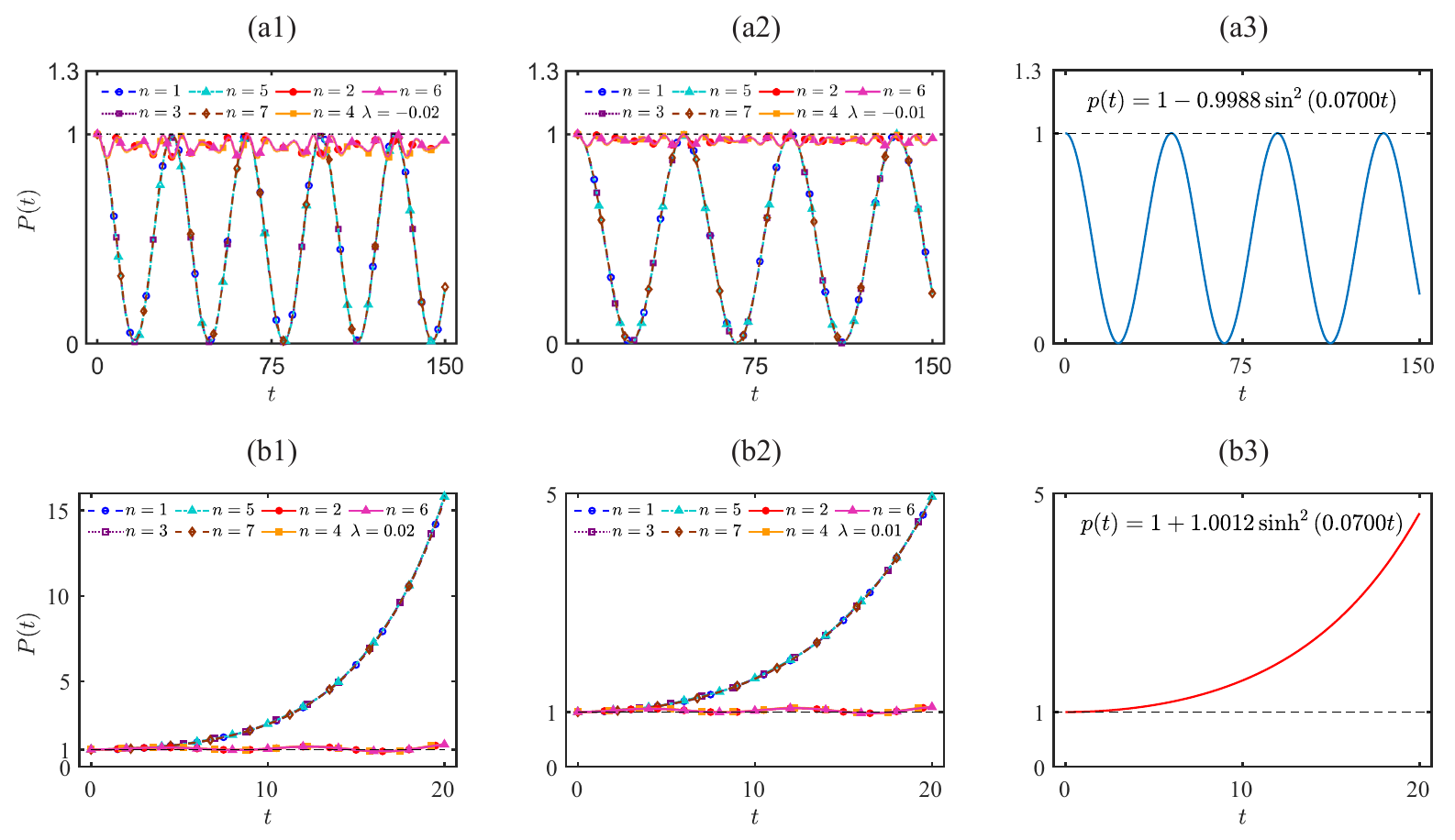}
	\caption{Profiles of the Dirac probability of time evolution for the
		condensate state $\left\vert \protect\psi _{n}\right\rangle $ under the
		Hamiltonian $H_{\mathrm{Pos}}$, given by Eq. (\protect\ref{HpreHposHp}) with 
		$N=8$ and small $\protect\lambda $. (a1, a2) Plots of $P(t)$\ in Eq. (%
		\protect\ref{P(t)}) obtained by exact diagonalization for the Hamiltonians
		with $\protect\lambda =-0.02$ and $\protect\lambda =-0.01$, respectively.
		(b1, b2) The same plots for the cases with $\protect\lambda =0.02$ and $%
		\protect\lambda =0.01$ respectively. For comparison, we also plot $P(t)$ in
		(a3) and (b3), given by Eqs. (\protect\ref{p1}) and (\protect\ref{p2}) with $%
		\protect\gamma =\pm 0.00245$, which are also presented in the panels,
		respectively. The plots of $P(t)$\ match our predictions: showing that the
		initial states with odd and even particle filling exhibit distinct dynamical
		behaviors. We can see that the curves in (a1, a2)\ and (b1, b2) are similar
		to those in (a3) and (b3), implying that the states $\left\vert \protect\psi %
		_{n}\right\rangle $\ are second-order coalescing states when $n$ is odd.}
	\label{fig2}
\end{figure*}

\section{Parity-dependent dynamics}

\label{Parity-dependent dynamics}

According to non-Hermitian quantum theory, the presence or absence of an EP
gives rise to distinct dynamical behaviors. Therefore, initial states with
different particle-number parity exhibit different dynamical behaviors under
the evolution governed by $H_{\mathrm{HB}}$. Specifically, the dynamics of
the system with parameters far away from, near, and at the EP exhibit
extremely different behaviors. To proceed, we first provide a brief review
of the time evolution of a two-level non-Hermitian system.

The matrix of the Hamiltonian is

\begin{equation}
h=\left( 
\begin{array}{cc}
i\gamma & 1 \\ 
1 & -i\gamma%
\end{array}%
\right) ,
\end{equation}%
which has the coalescing state 
\begin{equation}
\phi _{\text{\textrm{c}}}=\frac{1}{\sqrt{2}}\left( 
\begin{array}{c}
1 \\ 
-i%
\end{array}%
\right) ,
\end{equation}%
at the EP with $\gamma =1$. We are interested in the time evolution of the
coalescing state $h$\ under the Hamiltonian $h$\ with different values of $%
\gamma $. The evolved state reads $\phi (t)=e_{\text{\textrm{c}}}^{-iht}\phi 
$, and its Dirac probability $P(t)=\left\vert \phi (t)\right\vert ^{2}$\ can
be obtained as following.

(i) In the case where $\gamma <1$, the Dirac probability can be written as%
\begin{equation}
p(t)=1-\frac{2\gamma }{\gamma +1}\sin ^{2}(\sqrt{1-\gamma ^{2}}t),
\label{p1}
\end{equation}%
which is a periodic function with period $\pi /\sqrt{1-\gamma ^{2}}$. Its
minimum value is $1-2\gamma /(\gamma +1)$, which approaches zero when the
system is in the vicinity of the EP. In contrast, $p(t)$ remains at $1$ when 
$\gamma $ is near zero.

(ii) In the case where $\gamma >1$, the Dirac probability can be written as%
\begin{eqnarray}
p(t) &=&1+\frac{2\gamma }{\gamma +1}\sinh ^{2}\left( \sqrt{\gamma ^{2}-1}%
t\right)  \notag \\
&\approx &\frac{\gamma }{2\left( \gamma +1\right) }e^{2\sqrt{\gamma ^{2}-1}%
t},  \label{p2}
\end{eqnarray}%
which is an exponential function.

Inspired by the above example, we investigate the difference between the
even and odd particle-number fillings for $H_{\mathrm{HB}}$ by conducting
numerical simulations of the subsequent quenching process. The corresponding
Hamiltonians are defined as%
\begin{eqnarray}
H_{\mathrm{Pre}} &=&H_{\mathrm{HB}},  \notag \\
H_{\mathrm{Pos}} &=&H_{\mathrm{HB}}+i\lambda \left( a_{1}^{\dagger
}a_{1}-a_{N}^{\dagger }a_{N}\right) .  \label{HpreHposHp}
\end{eqnarray}%
The initial state $\left\vert \Phi (0)\right\rangle $ is taken as the
condensate state $\left\vert \psi _{m}\right\rangle $, given by Eq. (\ref%
{condensate state}), of the prequench Hamiltonian $H_{\mathrm{HB}}$.
To capture the effect of perturbation $H_{\mathrm{Pos}}-H_{\mathrm{Pre}}$ on
the dynamics, we compute the time evolution and measure the Dirac
probability of the evolved state $\left\vert \Phi (t)\right\rangle =e^{-iH_{%
\mathrm{Pos}}t}\left\vert \Phi (0)\right\rangle $. Here, we only consider
the cases with $\left\vert \lambda \right\vert \ll 1$ as perturbations.
According to the above analysis, the probability 
\begin{equation}
P(t)=\left\vert \left\vert \Phi (t)\right\rangle \right\vert ^{2},
\label{P(t)}
\end{equation}%
will vary drastically for the initial state $\left\vert \psi
_{m}\right\rangle $\ with odd $m$, while it remains approximately unchanged
for even $m$, in the small $\left\vert \lambda \right\vert $\ regime.

To verify and demonstrate the above analysis for iterating systems,
numerical simulations are performed for finite-size lattices with several
typical parameters $\gamma $ by exact diagonalization. Note that all the
above analysis focuses on the case with infinite $U$ and $\gamma =1$. The
effect of the deviation of $\gamma $\ can be investigated through numerical
simulation. We compute the temporal evolution for a single type of initial
state $\left\vert \psi _{m}\right\rangle $\ with different $m$. We plot the
probability $P(t)$\ in Fig. \ref{fig2}. {These numerical results accord with
our above analysis for the system with infinite }${U}$: (i) For initial
states $\left\vert \psi _{m}\right\rangle $\ with odd $m$, the probability $%
P(t)$\ oscillates with an amplitude near $1$ and with minima near zero when $%
\lambda <0$. It increases exponentially when $\lambda >0$. (ii) For initial
states $\left\vert \psi _{m}\right\rangle $\ with even $m$, the probability $%
P(t)$\ oscillates with a small amplitude when $\lambda <0$. It oscillates\
with an exponentially\ increasing amplitude when $\lambda >0$. In addition,
we also plot the exact results for the two-level non-Hermitian system, given
by Eqs. (\ref{p1}) and (\ref{p2}), for comparison. The results indicate that
the quench dynamic behavior of a finite-sized system exhibits
characteristics similar to those of the two-level non-Hermitian system. This
implies that the condensate states $\left\vert \psi _{n}\right\rangle $\ are
second-order coalescing states when $n$\ is odd.

\section{Summary}

\label{Summary}

In summary, we have revealed the dynamic sensitivity to particle-number
parity induced by the EPs in the hardcore Bose-Hubbard chain with two
end-site $\mathcal{PT}$-symmetric imaginary potentials. We have shown that a
set of condensate eigenstates can be exactly constructed when the fields are
in resonance. Furthermore, we found that these eigenstates have a peculiar
feature: they are coalescing states arising from the EP when the particle
number is odd, while they are not when the particle number is even. This
leads to the strong dependence of the dynamic behaviors on particle-number
parity. Our findings may stimulate research on the dynamic sensitivity to
particle-number parity due to its potential application on quantum
information processing.

\section*{Acknowledgment}

This work was supported by the National Natural Science Foundation of China
(under Grant No. 12374461).

\section*{Data availability}

The data that support the findings of this article are openly
available \cite{liu_sun_2025_17035648}. 

\section*{Appendix}

\label{Appendix}

In this appendix, we will show the identity in the main text,%
\begin{equation}
\left\langle \varphi _{n}|\psi _{n}\right\rangle =\left\{ 
\begin{array}{cc}
0, & (n\in \text{\textrm{odd}}) \\ 
\text{\textrm{Nonzero}}, & (n\in \text{\textrm{even}})%
\end{array}%
\right. ,
\end{equation}%
given by Eq. (\ref{bi norm}). From the definition of $\left\vert \psi
_{n}\right\rangle $ and $\left\vert \varphi _{n}\right\rangle $, we have%
\begin{equation}
\left\langle \varphi _{n}|\psi _{n}\right\rangle =\frac{1}{C_{2N}^{n}}%
\sum_{\{l_{j}\}}(-1)^{\sum_{j=1}^{n}l_{j}},
\end{equation}%
where $\{l_{j}\}$\ denotes a combination, or a selection of $n$ distinct
numbers from the set $[1,2N]$. Here $\sum_{\{l_{j}\}}$ denotes the sum over
all combinations, and the total number of combinations is $C_{2N}^{n}$. For
each combination $\{l_{j}\}$, the number $\sum_{j=1}^{n}l_{j}$\ is either
even or odd. We define $s_{\text{\textrm{odd}}}\ $($s_{\text{\textrm{even}}}$%
) as the number of total combinations with odd (even) of the sum $%
\sum_{j=1}^{n}l_{j}$, which can be obtained as%
\begin{equation}
s_{\text{\textrm{even}}}=\sum_{i\in \text{\textrm{even}}%
}C_{N}^{i}C_{N}^{n-i},
\end{equation}%
and%
\begin{equation}
s_{\text{\textrm{odd}}}=\sum_{i\in \text{\textrm{odd}}}C_{N}^{i}C_{N}^{n-i}.
\end{equation}%
Then we have 
\begin{equation}
s_{\text{\textrm{even}}}+s_{\text{\textrm{odd}}}=C_{2N}^{n},
\end{equation}%
and%
\begin{equation}
s_{\text{\textrm{even}}}-s_{\text{\textrm{odd}}}=%
\sum_{i}(-1)^{i}C_{N}^{i}C_{N}^{n-i}.
\end{equation}%
We note that the number $s_{\text{\textrm{even}}}-s_{\text{\textrm{odd}}}$
is crucial to the value of $\left\langle \psi _{n}^{-}|\psi
_{n}^{+}\right\rangle $, that is%
\begin{equation}
\left\langle \varphi _{n}|\psi _{n}\right\rangle =\left\{ 
\begin{array}{cc}
\text{\textrm{Nonzero,}} & s_{\text{\textrm{even}}}\neq s_{\text{\textrm{odd}%
}} \\ 
0, & s_{\text{\textrm{even}}}=s_{\text{\textrm{odd}}}%
\end{array}%
\right. .
\end{equation}%
On the other hand, the quantity $\sum_{i}(-1)^{i}C_{N}^{i}C_{N}^{n-i}$\ can
be envalued from the generating function $(1-x)^{N}(1+x)^{N}=(1-x^{2})^{N}$.
It tells us the coefficient of $x^{n}$ is%
\begin{equation}
\sum_{i}(-1)^{i}C_{N}^{i}C_{N}^{n-i}=\left\{ 
\begin{array}{cc}
0\text{\textrm{,}} & (n\in \text{\textrm{odd}}) \\ 
(-1)^{n/2}C_{N}^{n/2}, & (n\in \text{\textrm{even}})%
\end{array}%
\right. .
\end{equation}%
Then the conclusion about $\left\langle \varphi _{n}|\psi _{n}\right\rangle $%
\ is approved.


\end{document}